Autor: Pablo Diniz Batista
\documentclass[10pt,aps,showkeys,twocolumn,article,groupedaddress]{revtex4-1}

\usepackage{graphicx}
\usepackage{dcolumn}
\usepackage{bm}
\usepackage{amssymb, amsmath}

\usepackage[T1]{fontenc}

\preprint{APS/123-QED}

\begin{document}

\title{An embedded measurement system for electrical characterization of \\ EGFET as pH sensor}

\author{Pablo Diniz Batista}
\email[]{batista@cbpf.br}
\homepage{http://batistapd.com}
\affiliation{Brazilian Center for Physics Research \\ Rua Dr. Xavier Sigaud, 150 - Urca- Rio de Janeiro - RJ - Brazil - CEP: 22290-180}
\homepage{http://cbpf.br}

\date{\today}

\begin{abstract}

This work presents the development of an electronic system for the electrical characterization of pH sensors based on the extended gate field effect transistor (EGFET). We designed an electronic circuit with a microcontroller (PIC15F14K50) as the main component in order to provide two programmable outputs voltage as well as circuits to measure electric current and voltages. The instrument performance analysis was carried out using a glass electrode as a sensitive membrane for investigating the EGFET operation as pH sensor. The results show that the system is an alternative to commercial equipment for the electrical characterization of sensors based on field effect devices. In addition, some of the key features expected of this electronic module are: low cost, flexibility, portability and communication with a personal computer using a USB port.

\end{abstract}


\maketitle

\section{Introduction}

Ion-sensitive field effect transistor (ISFET) firstly presented by Bergveld at the University of Twente in 1968 as the first chemical sensor using a semiconductor device of small dimensions\cite{REF01}. Soon after that, but a little later, Matsuo from Tohoku University in Japan, upon returning from a vacation in Stanford, published results showing a new type of device, similar to ISFET, first in a Japanese journal, and later, in an international journal \cite{REF02,REF03,REF04}. Over the years different applications using the ISFET were performed \cite{REF04,REF05}. For example, in 1983, J Van De Spiegel et al. reported the operation of the first extended gate field effect transistor (EGFET) based on the same ISFET operation principle except that the gate was not made directly on the metal oxide semiconductor field effect transistor (MOSFET) structure. EGFET was initially proposed as an alternative to ISFET to the detection of multiple substances in a single device at the same time \cite{REF06}. 
After 17 years, Li-Te Yin and colleagues presented the scientific community with a new alternative the manufacturing
EGFET. Unlike J Van Der Spiegel, they developed the EGFET from the connection between a membrane sensitive to hydrogen ions and a commercial MOSFET (CD4007UB) \cite{REF07,REF08}. 
Since the publication of these works, several papers have been published exploring both the optimization of the pH sensor devices and applications involving biosensors \cite{REF09,REF10,REF11,REF12,REF13,REF14,REF15}. In addition to that, it is worth stressing the scientific and technological development obtained with instruments involving these devices \cite{REF16,REF17,REF18,REF19,REF20,REF21,REF22}. 
Contributing to this perspective, this work aims at designing and building a dedicated electronic module for the electrical characterization of EGFET as pH sensor. Generally speaking, the procedures used to perform the electrical characterization of EGFET as a pH sensor are similar to those used to obtain the characteristic curves of the MOSFET except for the fact that, for the EGFET, the current is investigated as a function of the concentration of hydrogen ions present in the solution in which the membrane is immersed as shown in figure \ref{fig:Figure01}-a. So, in a nutshell, the measurement system must be capable of applying voltages to the MOSFET to the reference electrode whilst the current between the drain and source is monitored in such way that the value of these voltages as well as the current measurements can be remotely controlled by a personal computer allowing automation of experiments. Some of the key features expected of this electronic module are: low cost, flexibility, portability and communication with a personal computer using a USB port. 

This paper is organized as follows: In the second section we present an overview of both the hardware, consisting of analog and digital circuits, as well as the programs developed in C language. Then, in the third section, the results will be discussed with the object of analyzing system performance when it is used for the electrical characterization of EGFET as pH sensor considering a glass electrode as a selective membrane. Finally, we will present the conclusions and future prospects.

\section{Design and development}

\begin{figure}[htb]
\centering
\includegraphics[scale=0.40]{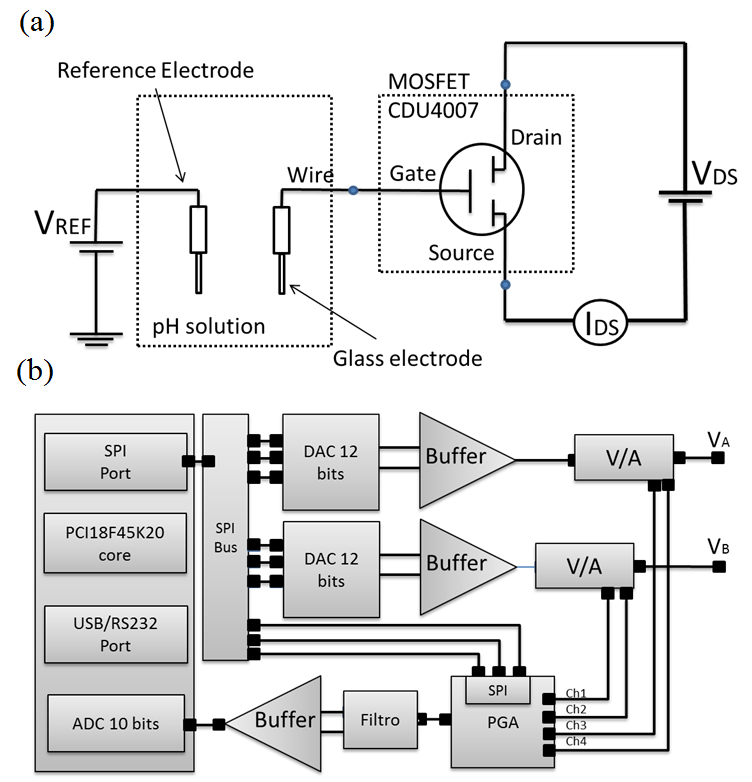}
\caption{(a) The EGFET as pH sensor consists of a sensitive membrane connected to a commercial MOSFET and a reference electrode. (b) The block diagram of the electronic module based on the microcontroller PIC18F45K20. It provides two programmable voltage outputs, voltmeters and ammeters.}
\label{fig:Figure01}
\end{figure}

\begin{figure}[htb]
\centering
\includegraphics[scale=0.7]{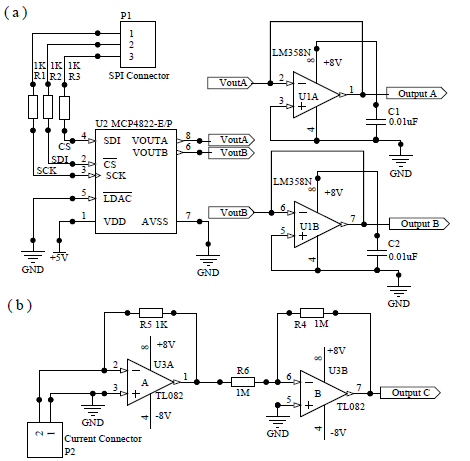}
\caption{ (a) Programmable voltage source electrical schematic based on the MCP4822. (b) Transimpedance circuit used to convert the electrical current between the source and drain of the MOSFET into a voltage.}
\label{fig:Figure02}
\end{figure}

\begin{figure*}[htb]
\centering                
\includegraphics[scale=0.50]{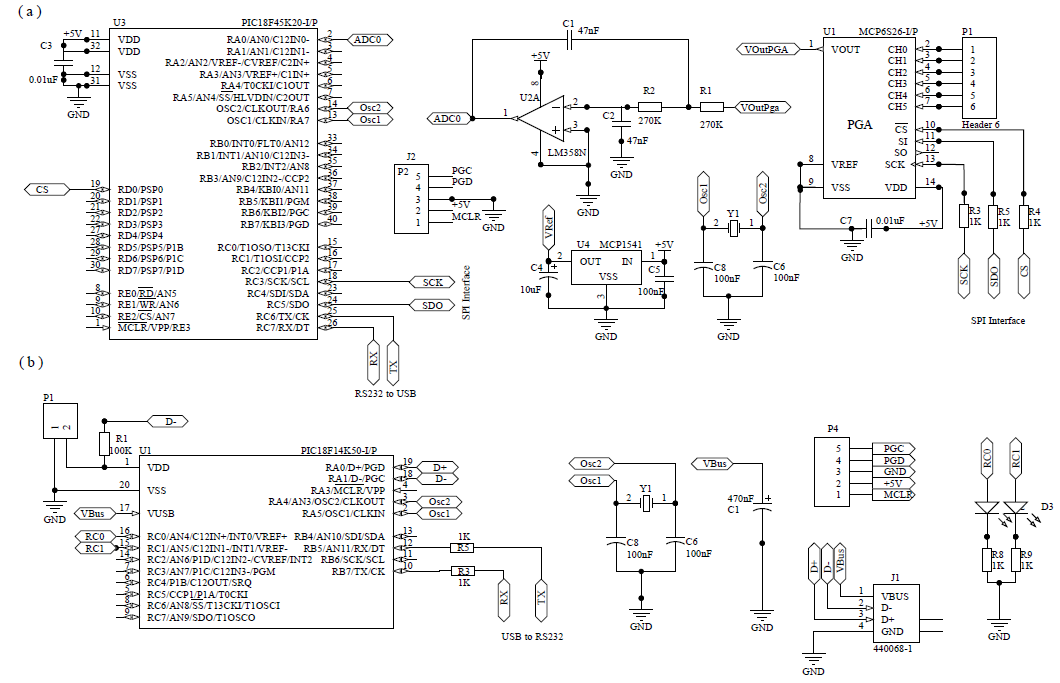}
\caption{ (a) Data acquisition system built with the microcontroller PIC18F14K50 wherein the ADC input is connected to a programmable gain operational amplifier MCP6S26 through a Sallen Key low-pass filter. (b) USB-RS232 converter electrical schematic using a PIC18F14K50.}
\label{fig:Figure03}
\end{figure*}

Figure \ref{fig:Figure01}-b shows the block diagrams for the measurement system. There is a digital-to-analog converter (DAC) connected to a voltage follower to be used as a programmable voltage source. For each voltage source there is a block dedicated to the voltmeter and ammeter whose outputs are multiplexed to an analog-to-digital converter (ADC) in the microcontroller via the programmable gain amplifier (PGA) of six inputs. Communication between the microcontroller and the peripherals is performed by the SPI protocol, whereas with the personal computer it is carried through a USB port. 
Figure \ref{fig:Figure02}-A presents the schematic diagram for a programmable voltage source with two outputs based on the MCP4822. Coupled to each output of this circuit is a voltage follower using the LM358 operational amplifier. The MCP4822 device is a DAC containing two outputs with 12-bit resolution and a serial communication according to the SPI protocol. Each of the two channels can operate either in active or off mode according to the values ​​present in the configuration registers. Then, the outputs A and B are monitored by the voltmeter and made available to the EGFET polarization with a resolution of 1 mV.
As shown in figure \ref{fig:Figure02}-b, the electronic module uses a transimpedance circuit followed by a unit gain amplifier based on the TL082 (operational amplifiers with high input impedance) to perform the current measurement. This circuit's main feature is the conversion of low values ​​of current into a voltage. It is distinguished by low sensitivity to parasitic capacitance of the circuit and by effective control of the feedback circuit stability \cite{REF23}. A feedback resistor provides a gain in the current-voltage relationship of 1:1000, i.e, a current of 1 mA at the input ammeter provides at the output a voltage of 1V. As the non-inverting input is grounded, the current values ​​will then be perceived by the inverting input, and thus appear in the output voltage with reverse polarity. Therefore, a second amplifier is used to provide both the polarity inversion and the impedance matching between the ammeter and the next stage.
Figure \ref{fig:Figure03}-a shows the electrical schematic of the data acquisition system which main component a microcontroller PIC18F14K50. Although the microcontroller presents multiple ADC inputs with 10 bits of resolution, we use the PGA to maintain flexibility. In the future, this feature will enable th use of an external ADC of 12 bits. Furthermore, MCP6S26 is capable of multiplexing up to six input channels with a gain of +1 V/V to +32 V/V. A standard SPI serial interface is used for the reception of instructions from a controller. At the PGA output, the signal is fed to analogue-to-digital converter via a Sallen Key low-pass with a cutoff frequency of 10 Hz \cite{REF24}.
%
The ADC reference voltage is provided by the integrated circuit MCP1541, so that a voltage of 4.096V provides an ADC resolution of 4 mV. The MCP1541 input voltage is connected to the $V_{IN}$ input device in parallel with the ceramic capacitor in order to reject the input capacitor noise voltage in the range of approximately 1 to 2 MHz. Noise above 2 MHz is far beyond the bandwidth of the reference voltage, and hence will not be transmitted from the input pin to the output. The load capacitance is required to stabilize the reference voltage. Finally, a converter RS-232 to USB was developed in order to provide greater flexibility using a PIC18F14K50 microcontroller as shown in figure \ref{fig:Figure03}-b. 

To put this instrument into operation, it is still necessary to develop two programs, one for the microcontroller and the other for the computer. The main idea is that with this application we can access all the peripherals connected to the microcontroller. In other words, at this point the microcontroller can be thought simply as a bridge between the personal computer and the peripherals connected to the microcontroller. This architecture is being proposed in order to achieve a greater flexibility in the usage of the hardware, taking into account only the specific needs of each application. 
With this application it is possible, for instance, to set the voltages on channels 1 and 2 and to perform measurements of current and voltages across the ammeter and voltmeter in this module. In the first version, the program allows the selection of the initial and final voltage as well as the increment voltage for each channel. It is also possible to select how these values are varied over time. A trivial way to do so is to adjust the two sources at the same time so as one of them may or not remain at a constant value. The program has been developed to run on Windows operating system. However, the same source code can be easily adapted to other operating systems. The complete code for both programs as well as the electrical schematic can be freely downloaded at the following electronic address: http://batistapd.com.
\section{Results and discussion}
In this first analysis, the electronic module performance was investigated considering the operation of the voltmeter and the ammeter. Different calibration curves were obtained at the output of the MCP4822 using an Agilent digital multimeter connected to a personal computer via the USB port. The voltage is adjusted via the SPI port ​​between 0 and 4.095 V at intervals of 0.5 V. Furthermore, when the microcontroller receives a command to read data, it carries out the acquisition and conversion of the analog signal on the input of the ADC and then sends the results to the computer. The results, not shown here, demonstrated that the electronic module maintains the voltage set by the program over time confirming that the voltmeter can be used to monitor the outputs present in the programmable voltage source efficiently. Likewise, to test the ammeter performance different resistors were used as current loads and, based on the results we observe that the measuring current is satisfactory as expected, and thus, the ammeter can also be used to obtain the sensor curves.
\begin{figure}[htb]
\centering
\includegraphics[scale=0.17]{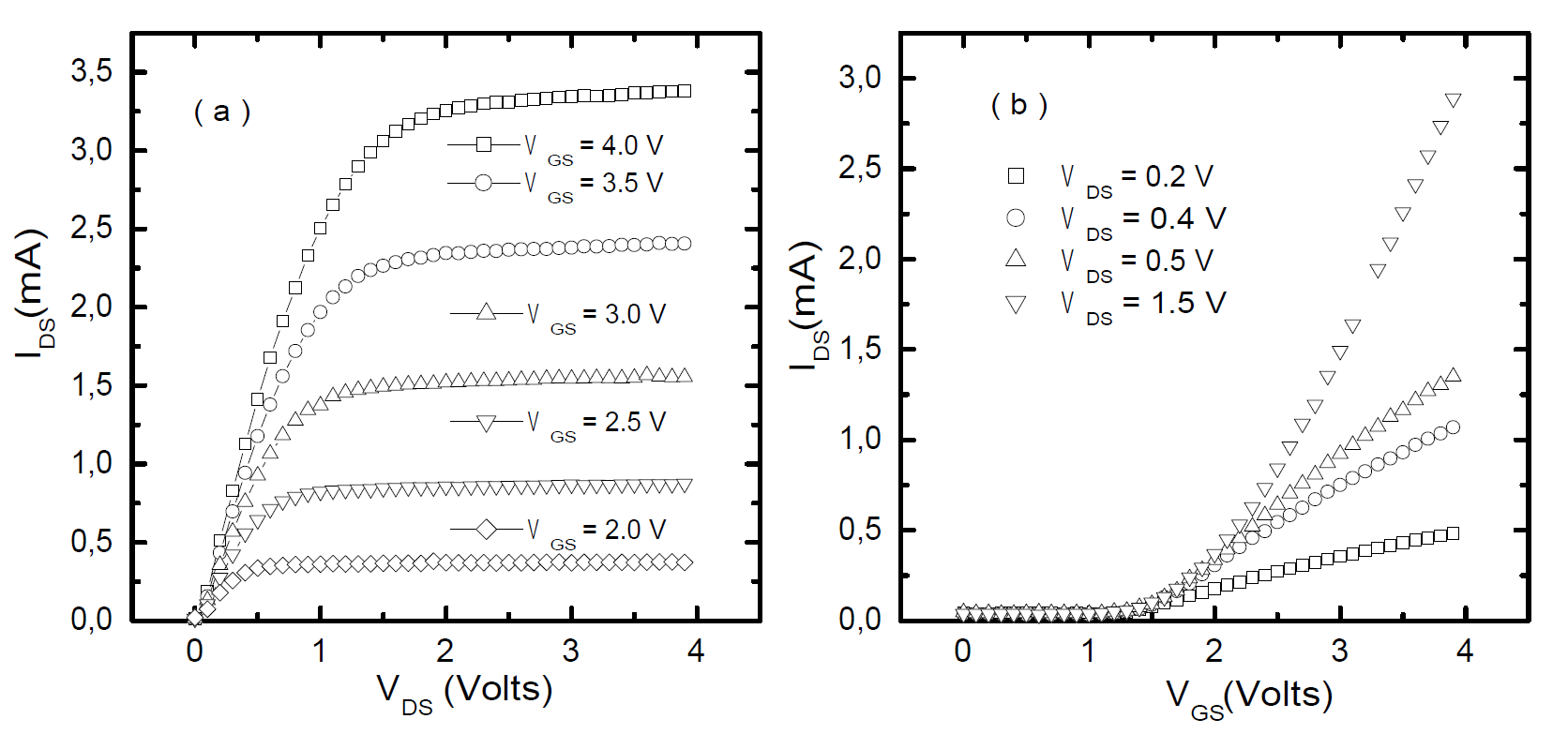}
\caption{ Electrical characterization of CD4007UB. (a) $I_{DS}$ versus $V_{DS}$ to different values of $V_{GS}$. (b) $I_{DS}$ versus $V_{GS}$ for different values of $V_{DS}$.}
\label{fig:Figure04}
\end{figure}
%
In the next step, as an application, this electronic module is used for the CD4007UB electrical characterization to determine the best condition for its operation as transducer. Figure \ref{fig:Figure04}-a shows the source-drain current $(I_{DS})$ as a function of the drain-source voltage $(V_{DS})$ for different values of gate voltages $(V_{GS})$. Note the presence of a saturation current whose amplitude is related to the gate voltage. The higher the value of $V_{GS}$, the larger the value of $I_{DS}$. Figure \ref{fig:Figure04}-b shows the $I_{DS}$ versus $V_{GS}$ curves for different values of $V_{DS}$. For low values of $V_{DS}$, i.e., $V_{DS} < 0.3 $ V, the MOSFET operates in the linear (Ohmic) region and the resistance is a function of $V_{DS}$. These data are meaninful to the computation of the device threshold voltage (about 1.5 V) and to EGFET sensitivity. 

As a criterion for comparison, the EGFET operation as pH sensor is investigated using this measurement system taking into account a glass electrode connected to the gate of the CD4007 as well as electrolyte buffers of known pH, more specifically $HCl$/$NaOH$ based solution with their pH value ranging between 2 and 12. This electrode is adapted to EGFET to evaluate the electronic module performance because the sensor has a sensitivity of 55 mV/pH in a range of 2 to 12. The wire related to pH signal is connected to the gate of the MOSFET, while the other connection (ground or reference voltage) remains disconnected from the circuit. To complete the measurement system the reference electrode is also immersed in the solution. In this case, the voltage previously applied to the gate is now applied to the reference electrode $(V_{REF})$.
\begin{figure}[htb]
\centering
\includegraphics[scale=0.24]{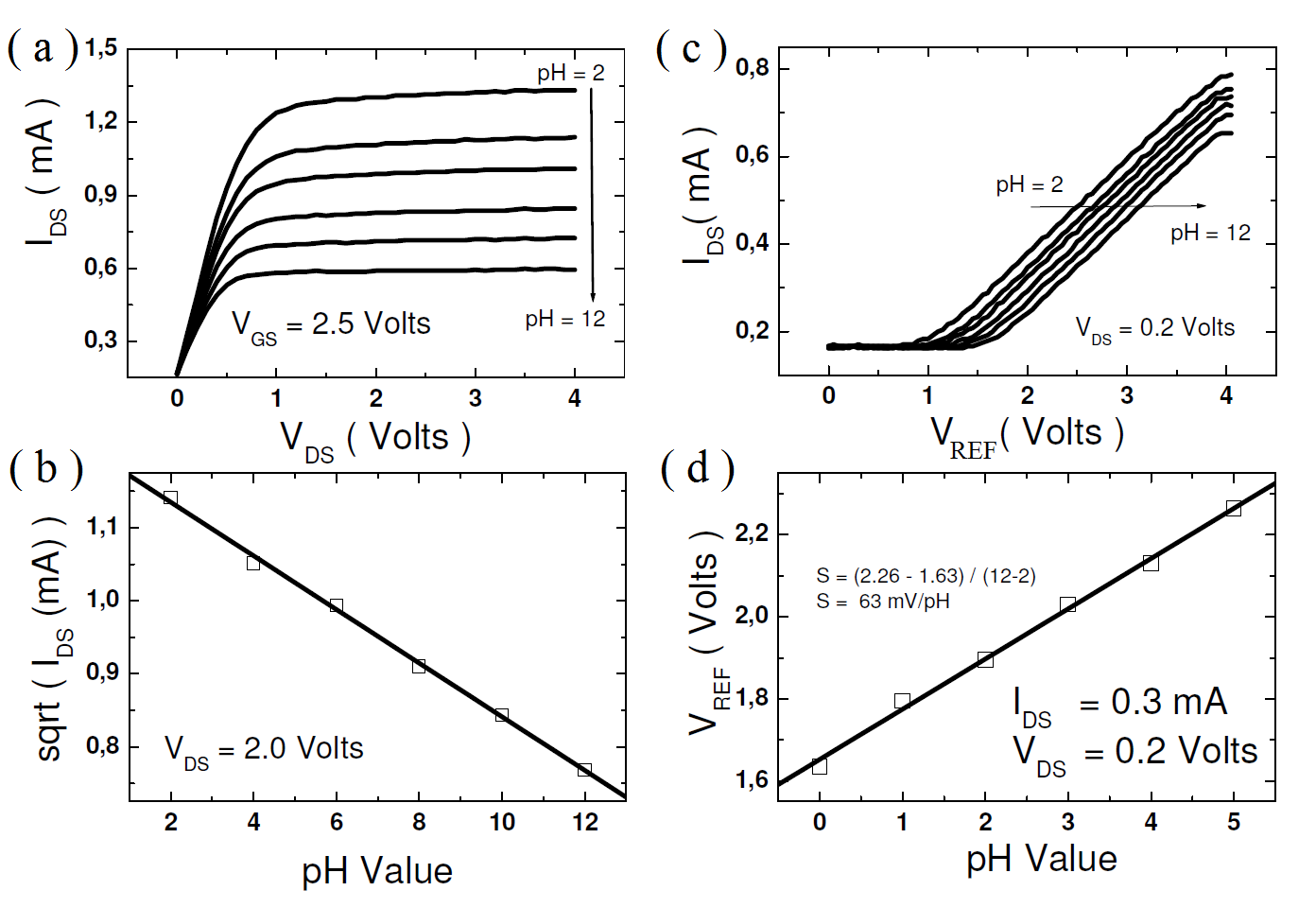}
\caption{ EGFET electrical characterization using a glass electrode connected to the gate of CD4007UB. (a) $I_{DS}$ versus $V_{DS}$for different pH values fixing $V_{REF}$ equal to 2.5 V. (b) Square root of $I_{DS}$ as a function of pH value for $V_{DS}$ equal to 2 V. (c) $I_{DS}$ versus $V_{REF}$ as a function of pH value considering $V_{DS}$ equal to 0.2 V. (d) EGFET sensitivity curve computed using $I_{DS}$ equal to 0.30 mA. Note that, the sensitivity of 63 mV/pH is higher than expected and it may be related to the temperature influence as well as with the precision of solutions used during the measurements. But best results will be achieved using a temperature control as well as standard solutions previously calibrated.
}

\label{fig:Figure05}
\end{figure}
Figure \ref{fig:Figure05}-a shows the EGFET characteristic curve considering the glass electrode immersed in solutions with different pH values. The current is measured by fixing $V_{REF}$ equal to 2.5 V. At the same time, $V_{DS}$ ranges from 0 to 4 V at intervals of 100 mV. The $I_{DS}$ depends on the pH value of the solution. Inasmuch as the pH varies from 2 to 12 the $I_{DS}$ decreases from 1.25 to 0.6 mA. From these results, we can observe a linear relation between the square root of $I_{DS}$ and the pH values as shown in figure \ref{fig:Figure05}-b.
Figure \ref{fig:Figure05}-c corresponds to the experiment wherein the current $I_{DS}$ is measured with $V_{DS}$ constant in a such way that the value of the EGFET sensitivity can be determined. Note that the current curves shift to the right as pH value varies from 2 to 12. It can be obtained plotting $V_{REF}$ as a function of pH for a current of 0.3 mA as shown in figure \ref{fig:Figure05}-d. 
%
\section{Conclusion}

This work aims to present this current stage both the design as well as developing a first prototype for scientific equipment as a potential tool to be used for research and technological development of pH sensors based on field effect transistor. Therefore, we focus our efforts to develop this prototype having microcontroller as the main component technology currently available on the market. For that, we design a low-cost easy-to-make electronic module to investigate the operation of the EGFET as a pH sensor. The sensor consists of nothing more than a membrane sensitive to hydrogen ions attached to the gate of a commercial MOSFET. Considering that the EGFET can be characterized electrically using the MOSFET traditional curves, the electronic module uses the PIC18F45K50 as a main component to provide two-channel programmable voltage as well as current and voltage meters. These peripherals are accessed by a program running on a personal computer via a USB communication in order to investigate the EGFET operation considering a glass electrode as a sensitive membrane. From the results presented, we can conclude that this measurement system is capable of generating a stable output voltage between 0 and 4.095 V. Finally, it was possible to show that the module obtains the characteristic curves of the EGFET functioning as the pH sensor satisfactorily. In addition to this feature, we were able to show that the developed instrument could also be an alternative to the ISFET electrical characterization as well as other types of sensors based on the field effect transistor. Eventually, this module will be coupled to a system for heating and temperature control of the solution, thus allowing a study of the influence of temperature on the operation of sensors.

\begin{acknowledgements}
 We are grateful to FAPERJ (E-26/110.997/2009) for funding this research project as well as Marcia Reis and Mônica Campiteli for reviewing the text.
\end{acknowledgements}


{}


\begin{thebibliography}{}

\bibitem{REF01} P. Bergveld, \emph{Development of an ion-sensitive solid-state device for neurophysiological measurements}, IEEE. Trans. Biomed. Eng. BME-17 (1970) 70-71.
\bibitem{REF02} T. Matsuo et al., \emph{Biomedical active electrode utilizing field effect of solid state device}, in Digest of Joing Meeting of Tohoku Sections of I.E.E.J. (1971) in Japanese.
\bibitem{REF03} T. Matsuo et al., \emph{An integrated field-effect electrode for biopotential recording}, IEEE. Trans. Biomed. Eng. BME-21 (1974) 485-487.
\bibitem{REF04} S. Middelhoek, \emph{Celebration of the tenth transducers conference: The past, present and future of transducer research and development}, Sensors and Actuators 82 (2000) 2-23.
\bibitem{REF05} M.J. Sh\"oning, \emph{Playing around with Field-Effect Sensors on the Basis of EIS Structures, LAPS and ISFET}, Sensors (2005), 5, 126-138.
\bibitem{REF06} J Van Der Spiegel et al., \emph{The extended gate chemically sensitive field effect transistor as multi-species microprobe}, Sensors and Actuators 4 (1983) 291-298.
\bibitem{REF07} Li-Te Yin et al., \emph{Separate structure extended gate $H^{+}$ ion sensitive field effect transistor on a glass substrate}, Sensor and Actuators B 71 (2000) 106-111.
\bibitem{REF08} Li-Lun Chi et al., \emph{Study on extended gate field effect transistor with tin oxide sensing membrane}, Materials Chemistry and Physics 63, Issue 1, (2000) 19-23.
\bibitem{REF09} Li-Lun Chi et al,. \emph{Study on separative structure of EnFet to detect acetylcholine}, Sensor and Actuators B: Chemical 71, Issues 1-2, (2000) 68-72.
\bibitem{REF10} Jai-Chyi Chen et al., \emph{Portable urea biosensor based on the extended-gate field effect transistor}, Sensors and Actuators B: Chemical 91 Issues 1-3 (2003) 180-186.
\bibitem{REF11} Li-Te Yin et al., \emph{Study of indium tin oxide thin film for separative extended gate ISFET}, Materials Chemistry and Physics 70 (2001) 12-16.
\bibitem{REF12} Jung-Chuan Chou et al., \emph{ $SnO_{2}$ Separative Structure Extended Gate $H^{+}$-Ion Sensitive Field Effect Transistor by Sol-Gel Technology and Readout Circuit Developed by Source Follower}, J. Appl. Phy. Vol 42 (2003) 6790-6794.
\bibitem{REF13} Batista P. D. et al., \emph{ZnO Extended-gate field-effect transistor as pH sensors}, Applied Physics Letter 87, 1435508 (2005).
\bibitem{REF14} Batista P. D. et al., \emph{ $SnO_{2}$ Extended Gate Field-Effect Transistor as pH sensor}, Brazilian Journal of Physics, vol 36, no 2A (2006).
\bibitem{REF15} Batista P. D. et al., \emph{Polycrystalline fluorine-doped tin oxide as sensoring thin film in EGFET pH sensor}, Journal Material of Science (2010) 45:5478-5481.
\bibitem{REF16} Wen-Yan Chung, \emph{ISFET interface circuit embedded with noise rejection capability}, Electronics Letters (2004) Vol. 40 No. 18 1115-1116.
\bibitem{REF17} Wen-Yan Chung et al., \emph{ISFET performance enhancement by using the improved circuit techniques}, Sensor and Actuators B 113 (2006) 555-562.
\bibitem{REF18} Wen-Yan Chung et al., \emph{New ISFET interface circuit design with temperature compensation}, Microelectronics Journal 37 (2006) 1105-1114.
\bibitem{REF19} D. Y. Chen et al., \emph{An Intelligent ISFET Sensory System with temperature and drift compensation for Long-term monitoring}, IEEE Sensors Journal, Vol. 8, No. 12, (2008)
\bibitem{REF20} Jung-Chuan Chou et al., \emph{Study on the Characteristics of the Measurement System for pH Array Sensors}, International Journal of Chemical and Biological Engineering 2:4 (2009).
\bibitem{REF21} Rani Rozina Abdul et al., \emph{Operation of ISFET as a pH sensor by using signal modulated reference electrode}, International Conference on Information and Multimedia Technology (2009) 548-550
\bibitem{REF22} Maciej Kokot et al., \emph{Excitation-independent constant conductance ISFET Driver}, Metrology and Measurement Systems Vol. XVI (2009), No 4, 631-640.
\bibitem{REF23} Keithley Instruments, \emph{Low Current Measurements}, Application Note Series (2007) Number 1671.
\bibitem{REF24} Texas Instruments, \emph{Analysis of the Sallen Key Architeture}, Application Report (2002).

\end{thebibliography}
\end{document}